\documentclass[graybox]{svmult}
\usepackage{graphicx}        
\usepackage{amsmath}
\usepackage{amssymb}

\DeclareMathSymbol{\lang}{\mathord}{symbols}{"68}
\DeclareMathSymbol{\rang}{\mathord}{symbols}{"69}
\DeclareMathSymbol{\openbra}{\mathord}{symbols}{"68}
\DeclareMathSymbol{\closeket}{\mathord}{symbols}{"69}
\newcommand{\ket}[1]{{| #1 \closeket}}

\begin{document}

\title*{Hamiltonian chaos with a cold atom in an optical lattice}
\author{S.V. Prants}
\institute{S.V. Prants \at Laboratory of Nonlinear Dynamical Systems,
Pacific Oceanological Institute of the Russian Academy of Sciences,
43 Baltiiskaya st., 690041 Vladivostok, Russia, \email{prants@poi.dvo.ru}}

\maketitle
\abstract{We consider a basic model of the lossless interaction between
a moving two-level atom and a standing-wave single-mode laser field.
Classical treatment of the translational atomic motion provides
the semiclassical Hamilton-Schr\"odinger equations of motion which are a
five-dimensional nonlinear dynamical system with two integrals of motion.
The atomic dynamics can be regular or
chaotic (in the sense of exponential sensitivity to small variations in
initial conditions and/or the system's control parameters) in dependence on
values of the control parameters, the atom-field detuning and recoil
frequency. We develop a semiclassical theory of
the chaotic atomic transport in terms of a random walk
of the atomic electric dipole moment $u$ which is one of the components
of a Bloch vector. Based on a jump-like behavior of
this variable for atoms crossing nodes of the standing
laser wave, we construct a stochastic map that
specifies the center-of-mass motion. We find the
relations between the detuning, recoil frequency and
the atomic energy, under which atoms may move in a rigid optical lattice
in a chaotic way. We obtain the analytical
conditions under which deterministic atomic transport has fractal properties
and explain a hierarchical structure of the dynamical fractals.
Quantum treatment of the atomic motion in a standing wave is studied in
the dressed state picture where the atom moves in two optical potentials
simultaneously. If the values of the detuning and a characteristic atomic
frequency are of the same order, than there is a probability of nonadiabatic
transitions of the atom upon crossing nodes of the standing wave.
At the same condition exactly, we observe sudden changes (jumps) in the
atomic dipole moment $u$ when the atom
crosses the nodes. Those jumps are accompanied  by splitting of
atomic wave packets at the nodes. Such a  proliferation of wave
packets at the nodes of a standing wave is a manifestation of classical
atomic chaotic transport. In particular, the effect of
simultaneous trapping of an atom in a well of one of the optical potential and its
flight in the other potential is a quantum analogue of a chaotic classical
walking of an atom. At large values of the
detuning, the quantum evolution is shown to be adiabatic in accordance with
a regular character of the classical atomic motion.}

\section{Short historical background}

The fundamental model for the interaction  of a radiation  with  matter,
comprising a collection of two-level quantum systems coupled with a single-mode
electromagnetic field,  provides the basis for laser physics and
describes a rich variety of nonlinear dynamical effects. The discovery
that a single-mode laser, a symbol of coherence and stability, may
exhibit deterministic instabilities and chaos is especially important
since lasers provide nearly ideal systems to test general ideas in
statistical physics. From the stand point of nonlinear dynamics,
laser is an open dissipative system which transforms an external
excitation into a coherent output in the presence of loss. In 1975 Haken
\cite{Haken} has shown that a single-mode, homogeneously broadened laser,
operating on resonance with the gain center can be described in the
rotating-wave approximation by three real semiclassical
Maxwell-Bloch equations which are isomorphic to the famous Lorenz equations.
Some manifestations of a Lorenz-type strange attractor and dissipative chaos
have been observed with different types of lasers.

In the same time George Zaslavsky with co-workers \cite{Zas} have studied
interaction of an ensemble of two-level atoms with their
own radiation field in a perfect single-mode cavity without any losses and
external excitations, which is known as the Dicke  model
\cite {Dicke}. They were able to demonstrate analytically and
numerically dynamical instabilities and
chaos of Hamiltonian type in a semiclassical version of the
Dicke  model without rotating-wave approximation.
It was the first paper that opened the door to study Hamiltonian
atomic chaos in the rapidly growing fields of cavity
quantum electrodynamics, quantum and atomic optics.
Semiclassical equations of motion for this system may be
reduced to Maxwell-Bloch equations for three real independent
variables which, in difference from the laser theory, do not include
losses and pump. Those equations are, in general, nonintegrable, but
they become integrable immediately after adopting the
rotating-wave approximation  \cite{Jaynes}
that implies the existence of an additional integral of  motion,
conservation of the so-called number of excitations.
Numerical experiments have shown that prominent chaos
arises when the density of atoms is very large (approximately
$10^{20}$ cm${}^3$ in the optical range \cite{Zas}). The following progress
in this field has been motivated, mainly, by a desire to find manifestations
of  Hamiltonian atomic chaos in the models more suitable for experimental
implementations. Twenty years after that pioneer paper,
manifestations of Hamiltonian chaos have been found in experiments with
kicked cold atoms in a modulated laser field. Nowdays, a few groups in
the USA, Australia, New Zealand, Germany, France, England, Italy and in other
countries can perform routine experiments on Hamiltonian chaos with cold
atoms in optical lattices and traps (for a review see \cite{Hens03}).

In this paper we review some results on theory of Hamiltonian chaos
with a single two-level atom in a standing-wave laser field that have
been obtained in our group in Vladivostok.
In spite of we published with George only one paper on this subject
\cite{PRA02}, our work in this field has been mainly inspired by his
paper \cite{Zas} written in 1975 in Krasnoyarsk, Siberia.

\section{Introduction}

An atom placed in a laser standing wave is acted upon by
two radiation forces, deterministic dipole and stochastic dissipative
ones \cite{Kaz}.
The mechanical action of light upon neutral atoms is at the
heart of laser cooling, trapping, and Bose-Einstein
condensation. Numerous applications of the mechanical action of light
include isotope separation, atomic lithography and epitaxy,
atomic-beam deflection and splitting, manipulating
translational and internal atomic states, measurement of atomic positions, and
many others. Atoms and ions in an optical lattice, formed by a laser standing wave, are perspective objects for
implementation of quantum information processing and
quantum computing. Advances in
cooling and trapping of atoms, tailoring optical potentials
of a desired form and dimension (including one-dimensional
optical lattices), controlling the level of dissipation
and noise are now enabling the direct experiments
with single atoms to study fundamental
 principles of quantum physics, quantum chaos, decoherence,
and quantum-classical correspondence (for recent reviews on cold atoms
in optical lattices see Ref.~\cite{GR01,MO06}).

Experimental study of quantum chaos
has been carried out with ultracold atoms  in $\delta$-kicked
optical lattices \cite{MR94,RB95,Hens03}.
To suppress spontaneous emission and provide a coherent quantum
dynamics atoms in those experiments were
 {\em detuned far
from the optical resonance}. Adiabatic elimination of
the excited state amplitude leads to an effective
Hamiltonian for the center-of-mass motion \cite{GSZ92},
whose 3/2 degree-of-freedom classical analogue
 has a mixed phase space with
regular islands embedded in a chaotic sea.
De Brogile waves of $\delta$-kicked ultracold atoms have been shown to
demonstrate under appropriate conditions the effect
of dynamical localization in momentum
distributions which means the quantum suppression of chaotic diffusion
\cite{MR94,RB95,Hens03}. Decoherence due to spontaneous emission or noise tend to suppress this
 quantum effect and restore classical-like dynamics. Another important
 quantum chaotic phenomenon with cold  atoms in far-detuned optical lattices
is a chaos-assisted tunneling. In experiments \cite{Steck01,HH01}
ultracold atoms have been
demonstrated to oscillate coherently between two regular
regions in mixed phase space even though the classical transport between
these regions is forbidden by a constant of motion (other than energy).

The transport of cold atoms in optical lattices
has been observed to take the form of ballistic motion, oscillations in wells of
the optical potential, Brownian motion \cite{Chu85}, anomalous diffusion
and L\'evy
flights \cite{BB02,ME96}. The L\'evy flights
have been found in the context of subrecoil laser cooling \cite{BB02} in the distributions of
escape times for ultracold atoms trapped in the potential wells with
momentum states close to the dark state. In those experiments the variance and
 the mean time for atoms to leave the trap have been shown
to be infinite.

A new arena of
quantum nonlinear dynamics with atoms in optical
lattices is opened if we work {\em near the optical resonance} and take the
 dynamics of
internal atomic states into account. A single atom in a standing-wave
laser field may be semiclassically treated as a nonlinear dynamical
system with coupled internal (electronic) and external (mechanical) degrees
of freedom \cite{PRA01,JETPL01,JETPL02}. In the semiclassical
and Hamiltonian limits (when one treats atoms as point-like particles and neglects
spontaneous emission and other losses of energy), a number of nonlinear dynamical
effects have been analytically and numerically demonstrated
with this system: chaotic Rabi oscillations
 \cite{PRA01,JETPL01,JETPL02}, Hamiltonian chaotic atomic
transport and dynamical fractals \cite{JETP03,PLA03,PRA07,PU06}, L\'evy flights and anomalous diffusion
\cite{PRA02,JETPL02,JRLR06}. These effects are caused by
local instability of the CM motion in a laser
field. A set of atomic trajectories under certain conditions becomes exponentially sensitive to small
variations in initial quantum internal and classical external states
or/and in the control parameters, mainly, the atom-laser detuning. Hamiltonian evolution
is a smooth
process that is well described in a semiclassical approximation by the coupled Hamilton-Schr\"odinger
equations. A detailed theory of Hamiltonian chaotic transport of
atoms in a laser standing wave has been developed in our recent paper
\cite{PRA07}.

\section{Semiclassical dynamics}
\subsection{Hamilton-Schr\"odinger equations of motion}

We consider a two-level atom with mass $m_a$ and transition
frequency $\omega_a$
in a one-dimensional classical standing laser wave with the frequency $\omega_f$
and the wave vector $k_f$. In the frame
rotating with the frequency $\omega_f$, the Hamiltonian is
the following:
\begin{equation}
\hat H=\frac{\hat P^2}{2m_a}+\frac{1}{2}\hbar(\omega_a-\omega_f)\hat\sigma_z-
\hbar \Omega\left(\hat\sigma_-+\hat\sigma_+\right)\cos{k_f \hat X}.
\label{Jaynes-Cum}
\end{equation}
Here $\hat\sigma_{\pm, z}$ are the Pauli operators which describe the transitions
between lower, $\ket{1}$, and upper, $\ket{2}$, atomic states,
$\Omega$ is a maximal value of  the Rabi frequency.
The laser wave is assumed to be strong enough,
so we can treat the field classically. Position $\hat X$ and momentum $\hat P$
operators will be considered in section ``Semiclassical dynamics''
as $c$-numbers, $X$ and $P$. The simple wavefunction for the electronic
degree of freedom is
\begin{equation}
\ket{\Psi(t)}=a(t)\ket{2}+b(t)\ket{1},
\label{Psi}
\end{equation}
where $a$ and $b$ are the complex-valued probability amplitudes to find the
atom in the states $\ket{2}$ and $\ket{1}$, respectively.
Using the Hamiltonian (\ref{Jaynes-Cum}), we get the Schr\"odinger equation
\begin{equation}
\begin{aligned}
i\frac{da}{dt}&=\frac{\omega_a-\omega_f}{2}a-\Omega b\cos k_fX,\\
i\frac{db}{dt}&=\frac{\omega_f-\omega_a}{2}b-\Omega a\cos k_fX.
\end{aligned}
\label{sysa}
\end{equation}
Let us introduce
instead of the complex-valued probability amplitudes $a$ and $b$ the
following real-valued variables:
\begin{equation}
u\equiv 2\operatorname{Re}\left(ab^*\right),\quad
v\equiv -2\operatorname{Im}\left(ab^*\right), \quad
z\equiv \left|a\right|^2-\left|b\right|^2,
\label{uvz_def}
\end{equation}
where $u$ and $v$ are a synchronized (with the laser field) and a quadrature
components of the atomic electric dipole moment, respectively, and $z$ is
the atomic population inversion.

In the process of emitting and
absorbing photons, atoms not only change their internal electronic states
but their external translational states change as well due to the photon
recoil. In this section we will describe the translational atomic motion
classically. The position and momentum of a point-like atom
satisfy classical Hamilton equations of motion. Full dynamics in the
absence of any losses is now governed by the Hamilton-Schr\"odinger
equations for the real-valued atomic variables
\begin{equation}
\begin{gathered}
\dot x=\omega_r p,\quad \dot p=- u\sin x, \quad \dot u=\Delta v,\\
\dot v=-\Delta u+2 z\cos x, \quad
\dot z=-2 v\cos x,
\end{gathered}
\label{mainsys}
\end{equation}
where $x\equiv k_f X$ and $p\equiv P/\hbar k_f$ are normalized
atomic center-of-mass position and momentum, respectively.
Dot denotes differentiation with respect to the dimensionless time $\tau\equiv \Omega t$.
The normalized recoil frequency, $\omega_r\equiv\hbar k_f^2/m_a\Omega\ll 1$,
and the atom-field detuning,
$\Delta\equiv(\omega_f-\omega_a)/\Omega$, are the control parameters.
The system has two integrals of motion, namely the total energy
\begin{equation}
H\equiv\frac{\omega_r}{2}p^2-u\cos x-\frac{\Delta}{2}z,
\label{H}
\end{equation}
and the Bloch vector $u^2+v^2+z^2=1$.
The conservation of the Bloch
vector length follows immediately from
Eqs. (\ref{uvz_def}).

\begin{figure}[htb]
\begin{center}
\includegraphics[width=0.45\textwidth,clip]{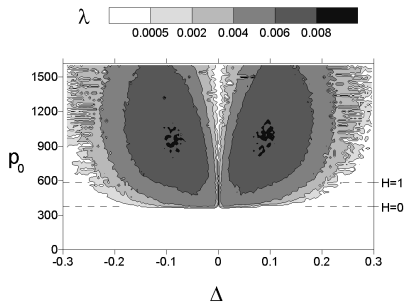}
\end{center}
\caption{Maximum Lyapunov exponent $\lambda$ vs atom-field detuning
$\Delta$ and initial atomic momentum $p_0$: $\omega_r=10^{-5}$, $u_0=z_0=0.7071$, $v_0=0$.}
\label{fig1}
\end{figure}

Equations (\ref{mainsys}) constitute a nonlinear Hamiltonian
autonomous system with two and half degrees of freedom which, owing to two integrals of motion, move on a three-dimensional
hypersurface with a given energy value $H$. In general, motion in a three-dimensional phase space in characterized by a positive
Lyapunov exponent $\lambda$, a negative exponent equal in magnitude to the positive one,
and zero exponent.
The maximum Lyapunov exponent characterizes the mean rate of the
exponential divergence of initially close trajectories and serves as a quantitative measure of dynamical chaos in the system.
The result of computation
of the maximum Lyapunov exponent in dependence on
the detuning $\Delta$ and the initial atomic momentum $p_0$ is shown in
Fig.~\ref{fig1}. Color in the plot codes the value of the maximum Lyapunov exponent $\lambda$.
In white regions the values of $\lambda$ are almost zero,
and the atomic motion is regular in the corresponding ranges
of $\Delta$ and $p_0$. In
shadowed regions positive values of $\lambda$ imply unstable motion.

Figure~\ref{fig1} demonstrates that the center-of-mass motion becomes unstable
if the dimensionless momentum exceeds the value $p_0 \approx 300$ that
corresponds (with our normalization) to the atomic velocity $v_a\approx 3$ m/s
for an atom with $m_a \approx 10^{-22}$~g in the field with the wavelength
close to the transition
wavelength $\lambda_a\simeq 800$~nm. With these estimates for the atomic and
lattice parameters and $\Omega/2\pi =10^9$, one gets the normalized
value of the recoil frequency equal to $\omega_r=10^{-5}$.
The detuning $\Delta$ will be varied
in a wide range, and the Bloch variables are restricted by the length
of the Bloch vector.

\subsection{Regimes of motion}

The case of exact resonance, $\Delta=0$, was considered in detail
in Ref. \cite{PRA01,JRLR06}.
Now we briefly repeat the simple results for the sake of self-consistency.
At zero detuning, the variable $u$ becomes a constant, $u=u_0$,
and the fast ($u$, $v$, $z$) and slow ($x$, $p$) variables are separated
allowing one to integrate exactly the reduced equations of motion.
The total energy (\ref{H}) is equal to $H_0=H(u=u_0, \Delta=0)$,
and the atom moves in a simple cosine potential $u_0 \cos x$
with three possible types of trajectories: oscillator-like motion
in a potential well if $H_0< u_0$
(atoms are trapped by the standing-wave field),
motion along the separatrix if $H_0= u_0$, and ballistic-like motion
if $H_0> u_0$.
The exact solution for the center-of-mass motion is easily found in
terms of elliptic functions (see \cite{PRA01,JRLR06}).

As to internal atomic evolution, it depends on the translational degree of freedom
since the strength of the atom-field coupling depends on the position of
atom in a periodic standing wave.
At $\Delta=0$, it is easy to find the exact solutions of Eqs.~(\ref{mainsys})
\begin{equation}
\begin{aligned}
v(\tau)&=\pm\sqrt{1-u^2}\ \cos\left(2\int\limits_0^\tau \cos  x\, d\tau'+
\chi_0\right),\\
z(\tau)&=\mp\sqrt{1-u^2}\ \sin\left(2\int\limits_0^\tau \cos  x\, d\tau'+
\chi_0\right),
\label{vz}
\end{aligned}
\end{equation}
where $u=u_0$, and $\cos[ x(\tau)]$ is a given function of
the translational variables only which can be found with the help of
the exact solution for $x$ \cite{PRA01,JRLR06}. The sign of $v$ is equal
to that for the initial
value $z_0$ and $\chi_0$
is an integration constant. The internal energy
of the atom, $z$, and its quadrature dipole-moment component $v$
could be considered as frequency-modulated signals
with the instant frequency $2\cos[ x(\tau)]$ and the modulation frequency
$\omega_r p(\tau)$,
but it is correct only if the maximum value of the first frequency is much greater than the value of the second one,
i.~e., for $|\omega_r p_0|\ll 2$.

\begin{figure}[htb]
\begin{center}
\includegraphics[width=0.45\textwidth,clip]{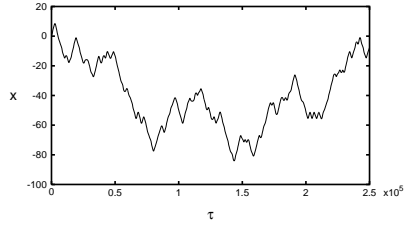}
\end{center}
\caption{Typical atomic trajectory in the regime of chaotic
transport: $x_0=0$, $p_0=300$, $z_0=-1$, $u_0=v_0=0$, $\omega_r=10^{-5}$,
$\Delta=-0.05$.}
\label{fig2}
\end{figure}

The maximum Lyapunov exponent $\lambda$ depends both on
the parameters $\omega_r$ and $\Delta$, and on initial conditions
of the system (\ref{mainsys}). It is naturally to expect
that off the  resonance atoms with comparatively small values
of the initial momentum $p_0$ will be at once trapped in the first well
of the optical potential, whereas those with large values of $p_0$ will
fly through. The question is what will happen with
atoms, if their initial kinetic energy will be close to the maximum of the optical potential.
Numerical experiments demonstrate that such atoms will wander in the
optical lattice with alternating trappings in the wells of the optical
potential and flights over its hills. The direction of the center-of-mass
motion of wandering atoms may change in a chaotic way (in the sense of exponential
sensitivity to small variations in initial conditions).
A typical chaotically wandering atomic trajectory is shown in Fig.~\ref{fig2}.

It follows from (\ref{mainsys}) that the translational motion of the atom
at $\Delta \neq 0$ is described by the equation of a nonlinear physical
pendulum with the frequency modulation
\begin{equation}
\ddot x+\omega_r  u(\tau)\sin x=0,
\label{12}
\end{equation}
where $u$ is a function of all the other dynamical variables.

\subsection{Stochastic map for chaotic atomic transport}

Chaotic atomic transport occurs even if the normalized
detuning is very small, $|\Delta|\ll 1$ (Fig.~\ref{fig1}).
Under this condition, we will derive in this section
approximate equations for the center-of-mass motion. The atomic
energy at $|\Delta|\ll 1$ is given with a good accuracy by its
resonant value $H_0$. Returning to the basic set of the equations of motion
(\ref{mainsys}),
we may neglect the first right-hand term in the fourth equation since it is
very small
as compared with the second one there. However, we cannot now exclude
the third equation from the consideration. Using the solution
(\ref{vz}) for $v$, we can transform this equation as
\begin{equation}
\dot u=\pm\Delta\sqrt{1-u^2}\ \cos\chi, \quad
\chi\equiv 2\int\limits_0^\tau \cos  x\, d\tau'+ \chi_0.
\label{u}
\end{equation}
Far from the nodes of the standing wave, Eq. (\ref{u}) can be approximately
integrated under the additional condition,
$|\omega_r p|\ll 1$, which is valid for the ranges of the parameters and the
initial atomic momentum where chaotic transport occurs. Assuming $\cos x$
to be a slowly-varying function in comparison with the
function $\cos\chi$, we obtain far from the nodes the approximate solution
for the $u$-component of the atomic dipole moment
\begin{equation}
u\approx\sin\left(\pm\frac{\Delta}{2\cos x} \sin \chi+C\right),
\end{equation}
where $C$ is an integration constant. Therefore, the amplitude
of oscillations of the quantity $u$ for comparatively slow
atoms ($|\omega_r p|\ll 1$) is small and of the order of $|\Delta|$
far from the nodes.

\begin{figure}[htb]
\begin{center}
\includegraphics[width=0.45\textwidth,clip]{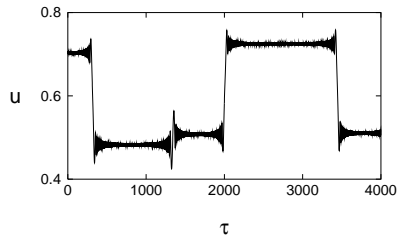}
\end{center}
\caption{Typical evolution of the atomic dipole-moment component
$u$ for a comparatively slow and slightly detuned atom:
$x_0=0$, $p_0=550$, $v_0=0$, $u_0=z_0=0.7071$,
$\omega_r=10^{-5}$, $\Delta=-0.01$.}
\label{fig3}
\end{figure}

At $| \Delta | =0$, the synchronized component of the atomic dipole moment
$u$ is a constant whereas the other Bloch variables
$z$ and $v$ oscillate in accordance with the solution (\ref{vz}).
At $|\Delta | \neq 0$ and far from the nodes, the variable $u$ performs
shallow oscillations for the natural frequency $| \Delta |$ is small as
compared with the Rabi frequency. However,
the behavior of $u$ is expected to be very special when
an atom approaches to any node of the standing wave since near the node
the oscillations of the atomic population inversion $z$ slow down and
the corresponding driving frequency becomes close to the resonance with the
natural frequency. As a result,
sudden ``jumps'' of the variable $u$ are expected to occur near the nodes.
This conjecture is supported by the numerical simulation. In
Fig.~\ref{fig3} we show a typical behavior of the variable $u$ for a
comparatively slow and slightly detuned atom.
The plot clearly demonstrates sudden ``jumps'' of $u$ near
the nodes of the standing wave and small oscillations between the nodes.

Approximating the variable $u$ between the nodes by
constant values, we can construct a discrete mapping \cite{PRA07}

\begin{equation}
u_{m}=\sin (\Theta \sin\phi_{m}+\arcsin u_{m-1}),
\label{u_m}
\end{equation}
where $\Theta \equiv |\Delta|\sqrt{\pi/\omega_r p_\text{node}}$ will be called an
angular amplitude of the jump, $u_m$ is a value of $u$ just after
the $m$-th node crossing,
$\phi_m$ are random phases to be chosen in the range
$[0,2\pi]$, and $p_\text{node}\equiv \sqrt{2H/\omega_r}$
is the value of the atomic momentum at the instant when the atom crosses
a node (which is the same with a given value of the energy $H$
for all the nodes).  With given values of $\Delta$,
$\omega_r$, and $p_\text{node}$, the map (\ref{u_m}) has been shown
numerically to give a satisfactory probabilistic distribution of
magnitudes of changes in the variable $u$ just after crossing
the nodes. The stochastic map (\ref{u_m}) is valid under the assumptions
of small detunings ($|\Delta|\ll 1$) and comparatively slow atoms
($|\omega_r p|\ll 1$). Furthermore, it is valid only for those ranges
of the control parameters and initial conditions where the motion
of the basic system (\ref{mainsys}) is unstable. For example,
in those ranges where all the Lyapunov exponents are zero, $u$ becomes
a quasi-periodic function and cannot be approximated by the map.

\subsection{Statistical properties of chaotic transport}

With given values of the control parameters and the energy $H$,
the center-of-mass motion is determined by the values of $u_m$
(see Eq. (\ref{12})). One can obtain from the expression for the
energy (\ref{H}) the conditions under which atoms continue to
move in the same direction after crossing a node
or change the direction of motion not reaching the
nearest antinode. Moreover, as in the resonance case,
there exist atomic trajectories along which atoms
move to antinodes with the velocity going
asymptotically to zero. It is a kind of separatrix-like motion with
an infinite time of reaching the stationary points.

The conditions for different regimes of motion depend on
whether the crossing number $m$ is even or odd.
Motion in the same direction occurs at $(-1)^{m+1}u_m<H$,
separatrix-like motion~---
at $(-1)^{m+1}u_m=H$, and turns~--- at $(-1)^{m+1}u_m>H$.
It is so because even values of $m$ correspond to $\cos x>0$,
whereas odd values~--- to $\cos x<0$. The quantity $u$
during the motion changes its values in a random-like manner
(see Fig.~\ref{fig3})
taking the values which provide the atom either to prolong the motion in
the same direction or to turn. Therefore, atoms may move chaotically
in the optical lattice. The chaotic transport occurs
if the atomic energy is in the range $0<H<1$.
At $H<0$, atoms cannot reach even the nearest node and oscillate
in the first potential well in a regular manner (see Fig.~\ref{fig1}).
At $H>1$, the values of $u$ are always satisfy to the flight condition.
Since the atomic energy is positive in the regime of
chaotic transport, the corresponding conditions can be
summarized as follows: at $|u|<H$,
atom always moves in the same direction, whereas at $|u|>H$,
atom either moves in the same direction, or turns depending on the sign
of $\cos x$ in a given interval of motion. In
particular, if the modulus of $u$ is larger for a long time then the
energy value, then the atom
oscillates in a potential well crossing two times
each of two neighbor nodes in the cycle.

\begin{figure}[htb]
\begin{center}
\includegraphics[width=0.45\textwidth,clip]{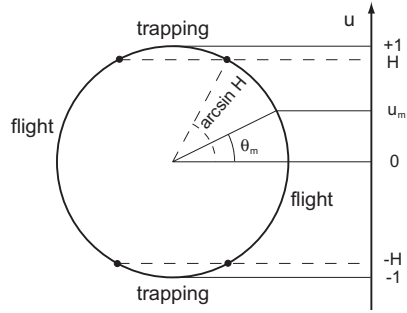}
\end{center}
\caption{Graphic representation for the maps of $u_m$ and
$\theta_m\equiv\arcsin u_m$. $H$ is a given value of the
atomic energy. Atoms either oscillate in optical potential
wells (trapping) or fly through the optical lattice (flight).}
\label{fig4}
\end{figure}

The conditions stated above allow to find a direct correspondence
between chaotic atomic transport in the optical
lattice and stochastic dynamics of the Bloch variable $u$.
It follows from Eq.~(\ref{u_m}) that the jump magnitude
$u_m-u_{m-1}$ just after crossing the $m$-th node depends
nonlinearly on the previous value $u_{m-1}$.
For analyzing statistical properties of the chaotic
atomic transport, it is more convenient to introduce
the map for $\arcsin u_{m}$~\cite{PRA07}
\begin{equation}
\theta_m\equiv\arcsin u_{m}= \Theta \sin\phi_{m}+\arcsin u_{m-1},
\label{u_ma}
\end{equation}
where the jump magnitude does not depend on a
current value of the variable. The map (\ref{u_ma}) visually
looks as a random motion of the point along a
circle of unit radius (Fig.~\ref{fig4}). The vertical projection
of this point is $u_m$. The value of the energy $H$ specifies
four regions, two of which correspond
to atomic oscillations in a well, and two other ones~---
to ballistic motion in the optical lattice.

We will call ``a flight'' such an event when atom passes, at least,
two successive antinodes (and three nodes). The continuous flight length $L>2\pi$ is
a distance between two successive turning points at which the atom changes
the sign of its velocity, and the discrete flight length is a number
of nodes $l$ the atom crossed. They are related in a simple way,
$L\simeq\pi l $, for sufficiently long flight.

Center-of-mass oscillations in a well of the
optical potential will be called ``a trapping''.
At extremely small values of the detuning, the
jump magnitudes are small and the trapping occurs,
largely, in the $2\pi$-wide wells, i.~e., in the space
interval of the length $2\pi$. At intermediate
values of the detuning, it occurs, largely, in the $\pi$-wide
wells, i. e. in the space interval of the length $\pi$.
Far from the resonance, $|\Delta|\gtrsim 1$, trapping
occurs only in the $\pi$-wide wells. Just like to the
case of flights, the number of nodes $l$, atom
crossed being trapped in a well, is a discrete measure
of trapping.

The PDFs for the flight $P_\text{fl}(l)$ and
trapping $P_\text{tr}(l)$ events were analytically derived to be exponential in
a case of large jumps \cite{PRA07}. In a case of small jumps,
the kind of the statistics depends on additional conditions imposed
on the atomic and lattice parameters, and the distributions
$P_\text{fl}(l)$ and $P_\text{tr}(l)$
were analytically shown to be either practically
exponential or functions with
long power-law segments with the slope $-1.5$ but exponential ``tails''.
The comparison of the PDFs computed
with analytical formulas, the stochastic map, and
the basic equations of motion has shown a good agreement in different ranges
of the atomic and lattice parameters \cite{PRA07}.
We will use the results obtained to find the analytical conditions,
under which the
fractal properties of the chaotic atomic transport can be observed, and
to explain the structure of the corresponding dynamical fractals.

Since the period and amplitude of the optical potential and the atom-field
detuning can be modified in a controlled way, the transport exponents of
the flight and trapping distributions are not fixed but can be varied
continuously, allowing to explore different regimes of the atomic transport.
Our analytical and numerical results with the idealized system have shown
that deterministic atomic transport in an optical lattice cannot be just
classified as normal and anomalous one. We have found that the flight and
trapping PDFs may have long algebraically decaying segments and a short
exponential ``tail''. It means that in some ranges of the atomic and lattice
parameters numerical experiments reveal
anomalous transport with L\'evy flights. The
transport exponent equal to $-1.5$ means that the first, second, and the other
statistical moments are infinite for a reasonably long time. The corresponding atomic trajectories
computed for this time are self-similar and fractal. The total distance,
that the atom travels for the time when the flight PDF
 decays algebraically, is dominated by a single flight. However,
the asymptotic behavior is close to normal transport.
In other ranges of the atomic and lattice parameters, the transport is
practically normal both for short and long times.

\subsection{Dynamical fractals}

Various fractal-like structures may arise in chaotic
Hamiltonian systems \cite{Gas,Zas05}. In Ref.
\cite{PLA03,JETP03,JRLR06,PU06} we have found numerically fractal properties
of chaotic atomic transport in cavities and optical lattices.
In this section we apply the analytical results of the theory
of chaotic transport, developed in the preceding
sections, to find the conditions under which the dynamical fractals may arise.

We place atoms one by one at the point $ x_0=0$
with a fixed positive value of the momentum $p_0$ and compute the time $T$ when they cross one
of the nodes at $x=-\pi/2$ or $x=3\pi/2$. In these
numerical experiments we change the value of the
atom-field detuning $\Delta$ only. All the
initial conditions $p_0=200$, $z_0=-1$, $u_0=v_0=0$ and
the recoil frequency $\omega_r=10^{-5}$ are fixed.
\begin{figure}
\begin{center}
\includegraphics[width=0.48\textwidth,clip]{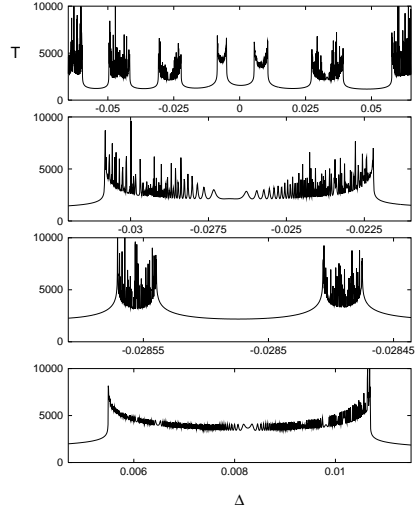}
\end{center}
\caption{Fractal-like dependence of the time of exit of atoms $T$
from a small region in the optical lattice on the detuning $\Delta$:
$p_0=200$, $z_0=-1$, $u_0=v_0=0$. Magnifications of the detuning
intervals are shown.}
\label{fig5}
\end{figure}
The exit time function $T(\Delta)$ in Fig.~\ref{fig5} demonstrates an intermittency of smooth
curves and complicated structures that cannot be resolved in principle, no
matter how large the magnification factor. The second and third panels in
Fig.~\ref{fig5}
demonstrate successive magnifications of the detuning intervals shown
 in the upper panel.
Further magnifications reveal a self-similar fractal-like structure that
is typical for Hamiltonian systems with chaotic scattering~\cite{Gas,BUP04}.
The exit time
$T$, corresponding to both the smooth and unresolved $\Delta$ intervals, increases
with increasing the magnification factor. Theoretically, there exist atoms
never crossing the border nodes at $x=-\pi/2$ or $x=3\pi/2$ in spite of the
fact that they have no obvious
energy restrictions to do that. Tiny interplay between chaotic external
and internal atomic dynamics prevents those atoms from leaving the small space
region.

Various kinds of atomic trajectories can be
characterized by the number of times $m$ atom crosses the central node at
$x=\pi/2$ between the border nodes.
There are also special separatrix-like
trajectories along which atoms asymptotically reach the points
with the maximum of the potential energy, having no more kinetic energy to
overcome it. In difference from the separatrix motion in the resonant
system ($\Delta=0$),  a detuned atom can
asymptotically reach one of the stationary points even if it was trapped for a while
in a well. Such an asymptotic motion
takes an infinite time, so the atom will never reach the border nodes.

The smooth $\Delta$ intervals in the first-order
structure (Fig.~\ref{fig5}, upper panel) correspond to atoms which never
change the direction of motion ($m=1$) and reach the border node at $x=3\pi/2$.
The singular
points in the first-order structure with $T=\infty$, which are located at the border between the
smooth and unresolved $\Delta$ intervals, are generated by the
asymptotic trajectories. Analogously, the smooth $\Delta$ intervals
in the second-order structure (second panel in Fig.~\ref{fig5}) correspond to
the $2$-nd order ($m=2$) trajectories, and so on.

The set of all the values of the detunings, generating
the separatrix-like trajectories, was shown to be a countable fractal in
Refs. \cite{JETP03,JRLR06},
whereas the set of the values generating dynamically trapped atoms with
$m=\infty$ seems to be uncountable. The exit time
$T$ depends  in a complicated way not only on the values of the
control parameters but on
initial conditions as well.

\begin{figure}
\begin{center}
\includegraphics[width=0.8\textwidth,clip]{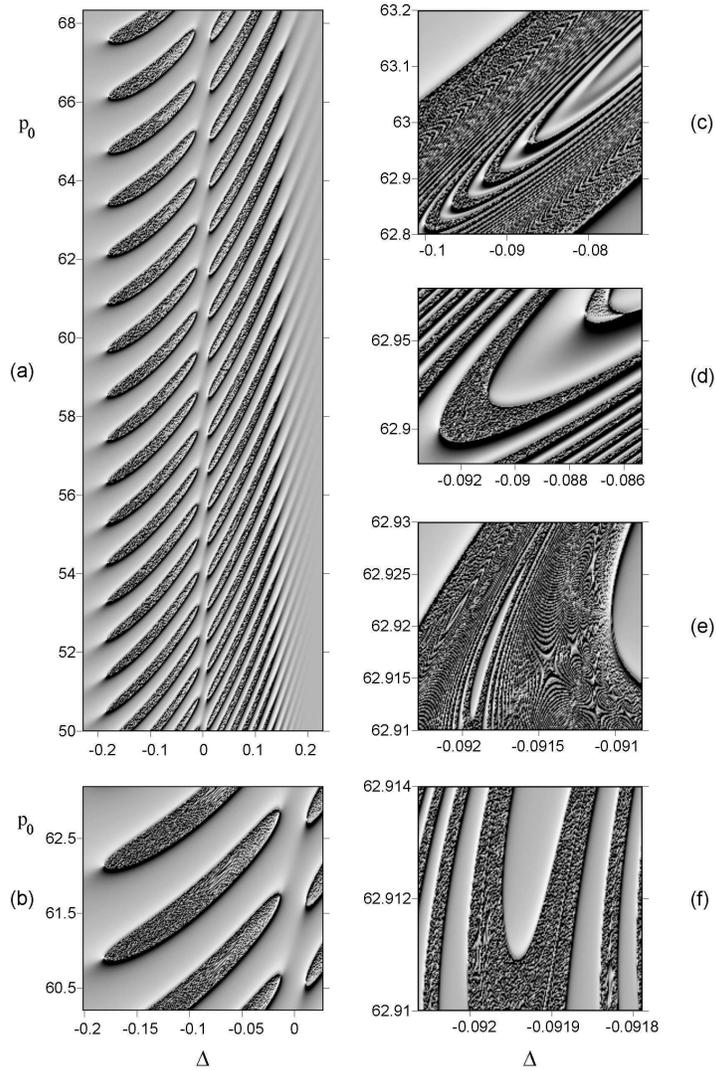}
\caption{The scattering function in the regime of chaotic wandering.
The time of exit $T$ vs the detuning $\Delta$
and the initial momentum $p_0$. The function is shown in a shaded
relief regime.}
\label{fig6}
\end{center}
\end{figure}

In Fig.~\ref{fig6} \cite{JRLR06} we presented a two-dimensional image
of the time of exit $T$  in the space of
the initial atomic momentum $p_0$ and the atom-field detuning $\Delta$.
A self-similarity of this function is evident.

The length of all smooth segments in the $m$-th order
structure in Fig.~\ref{fig5} is proportional to the number of atoms
$N(m)$ leaving the space $[-\pi/2, 3\pi/2]$ after crossing the central node
$m$ times. An exponential scaling $N(m) \sim \exp (-\gamma m)$ has been found
numerically with $\gamma \simeq 1$. The trapping PDFs, computed with the
basic and reduced equations of motion at the detunings in the range
shown in Fig.~\ref{fig5}, have been found to have exponential tails.
It is well known \cite{Gas} that Hamiltonian systems with fully
developed chaos demonstrate, as a rule,
exponential decay laws, whereas the systems with a mixed phase space
(containing islands of regular motion) usually have more slow algebraic decays
due to the effect of stickiness of trajectories to the boundaries of such
islands \cite{Zas05}. We have not found visible regular islands in our system
at the values of the control parameters used to compute the fractal
in Fig.~\ref{fig5} and we may conclude that the exponential scaling is a result
of completely chaotic wandering of atoms in the space interval
$[-\pi/2, 3\pi/2]$ resembling chaotic motion in hyperbolic systems.

The fractal-like structure with smooth and unresolved components
may appear if atoms have an alternative either to turn back or to prolong the
motion in
the same direction just after crossing the node
at $x=\pi/2$. For the first-order structure in
the upper panel in Fig.~\ref{fig5}, it means that the internal
variable $u$ of an atom, just after crossing the node
for the first time ($\cos x<0$), satisfies either to the condition $u_1<H$
(atom moves in the same direction), or to the condition $u_1>H$
(atom turns back). If $u_1=H$, then the exit time $T$ is
infinite. The jumps of the variable $u$ after crossing the node
are deterministic but sensitively dependent
on the values of the control parameters and initial
conditions. We have used this fact when introducing the stochastic map.
Small variations in these values lead to
oscillations of the quantity $\arcsin u_1$ around the initial
value $\arcsin u_0$ with the angular
amplitude $\Theta$. If this amplitude is large
enough, then the sign of the quantity $u_1-H$ alternates and we obtain alternating
smooth (atoms reach the border $x=3\pi/2$ without changing their direction of
motion) and unresolved (atoms turns a number of times
before exit) components of the fractal-like structure.

If the values of the parameters admit large jump
magnitudes of the variable $u$, then
the dynamical fractal arises in the energy range $0<H<1$, i.~e., at the
same condition under which atoms move in the optical lattice
in a chaotic way. In a case of small jump magnitudes,
fractals may arise if the initial value of an atom $u_0$ is close
enough to the value of the energy $H$, i.~e., the atom has
a possibility to overcome the value $u=H$ in a single jump. Therefore,
the condition for appearing in the fractal $T(\Delta)$ the first-order
structure with singularities is the following:
\begin{equation}
|\arcsin u_0-\arcsin H|< \Theta.
\label{cond1}
\end{equation}

The generation of the second-order structure is explained analogously.
If an atom made a turn
after crossing the node for the first time, then it will cross the node for
the second time. After that, the atom either will turn or cross the border
node at $x=-\pi/2$. What will happen depend on the value of $u_2$.
However, in difference from the case with $m=1$, the condition for appearing
an infinite exit time with $m=2$ is $u_2=-H$.
Furthermore, the previous value $u_1$ is not fixed  (in difference from $u_0$)
but depends on the value of the detuning $\Delta$. In any case we have
$u_1>H$ since the second-order structure consists of the trajectories of those
atoms which turned after the first node crossing. In order for an atom would be able to turn
after the second node crossing, the magnitude of its variable
$u$ should change sufficiently to be in the range $u_2<-H$. The atoms, whose
variables $u$ could not ``jump'' so far, leave the space $[-\pi/2, 3\pi/2]$.
The singularities are absent in the middle segment of
the second-order structure shown in the second panel in Fig.~\ref{fig5}
because all the corresponding atoms left the space after the second node crossing.
The variable $u_2$ oscillates with varying $\Delta$ generating oscillations
of the exit time. The condition for appearing singularities in
the second-order structure is the following:
\begin{equation}
2 \arcsin H< \Theta.
\label{cond2}
\end{equation}
With the values of the parameters taken in the simulation,
we get the energy $H=0.2+\Delta/2$. It is easy
to obtain from the inequality (\ref{cond2}) the approximate value of the
detuning $|\Delta|\approx 0.0107$ for which the second-order
singularities may appear. In the lower panel in Fig.~\ref{fig5}
one can see this effect. No additional conditions are required for
generating the structures of the third and the next orders.

Inequality (\ref{cond2}) is opposite to the inequality
that determines the condition for appearing power law decays in the flight
PDF. Therefore, dynamical fractal may appear in those ranges of
the control parameters where the L\'evy flights are impossible and vice versa.
However, the trapping PDF may have a power law decay.
Inequality (\ref{cond2}) in difference from (\ref{cond1}) is strongly
related with the chosen concrete scheme for computing exit times.
It is not required with other schemes, say, with three antinodes
between the border nodes.

\section{Quantum dynamics}

In this section we will treat atomic translational motion
quantum mechanically, i.~e., atom is supposed to be not a point particle
but a wave packet.
The corresponding Hamiltonian $\hat H$ has the form (\ref{Jaynes-Cum})
with $\hat X$ and  $\hat P$ being the position and momentum operators.
We will work in the momentum space with the state vector
\begin{equation}
\ket{\Psi(t)} = \int \left(a(P,t)\ket{2} + b(P,t)\ket{1}\right)\ket{P}dP,
\label{psip}
\end{equation}
which satisfies to the Schr\"odinger equation
\begin{equation}
i\hbar\frac{d\ket{\Psi}}{dt}=\hat H\ket{\Psi}.
\label{Sch}
\end{equation}
The normalized equations for the probability amplitudes have the form
\begin{equation}
\begin{aligned}
i \dot a(p)&= \frac12 (\omega_rp^2 - \Delta)a(p) -
\frac12[b(p+1) + b(p-1)], \\
i \dot b(p)&= \frac12 (\omega_rp^2 + \Delta)b(p) -
\frac12[a(p+1) + a(p-1)],
\label{Schp}
\end{aligned}
\end{equation}
with the same normalization and the control parameters as in the
semiclassical theory. When deriving (\ref{Schp}), we used
the following property of the momentum operator $\hat P$:
\begin{equation}
\cos k_f\hat X\ket{P}\equiv\frac12\left(e^{ik_f\hat X}+
e^{-ik_f\hat X}\right)\ket{P}=
\frac12\left(
\ket{P+\hbar k_f}+\ket{P-\hbar k_f}\right).
\label{oper}
\end{equation}
Equations (\ref{Schp}) are an
infinite-dimensional set of ordinary differential complex-valued
equations of the first
order with coupled amplitudes $a(p\pm n)$ and $b(p\pm m)$. To characterize
the internal atomic state, let us introduce the following variables;
\begin{equation}
\begin{aligned}
u(\tau)&\equiv 2 \operatorname{Re}\int dx \left[ a(x,\tau)b^*(x,\tau) \right], \\
v(\tau)&\equiv -2 \operatorname{Im} \int dx [a(x,\tau)b^*(x,\tau)], \\
z(\tau)&\equiv \int dx [|a(x,\tau)|^2 - |b(x,\tau)|^2],
\end{aligned}
\label{uvz}
\end{equation}
which are quantum versions of the Bloch components (\ref{uvz_def}), and we denote them
by the same letters.
\begin{figure}[htb]
\begin{center}
\includegraphics[width=0.45\textwidth,clip]{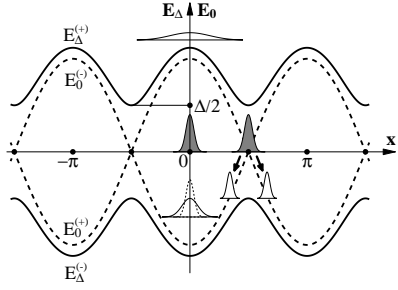}
\end{center}
\caption{Resonant $E_0^{(\pm)}$ and nonresonant $E_\Delta^{(\pm)}$ potentials
for an atom in a standing wave. The optical Stern-Gerlach effect in the
resonant potential is shown: splitting of an atomic wave packet launched at
the node of the wave ($x_0=\pi/2$, $p_0=0$). The wave packet, placed
initially at the antinode ($x_0=0$, $p_0=0$), appears to be simultaneously
at the top of $E_0^{(+)}$ and the bottom of $E_0^{(-)}$ potentials. Its
$\ket{+}$-component slides down both the sides of $E_0^{(+)}$ and the
$\ket{-}$-component oscillates at the bottom of $E_0^{(-)}$.
}
\label{fig7}
\end{figure}
\section{Dressed states picture and nonadiabatic transitions}

Interpretation of the atomic wave-packet motion in a standing-wave field is
greatly facilitated in the basis of atomic dressed states which are eigenstates
of a two-level atom in a laser field. The adiabatic dressed states
\begin{equation}
\begin{gathered}
\ket{+}_\Delta = \sin{\Theta}\ket{2} + \cos{\Theta}\ket{1}, \quad
\ket{-}_\Delta = \cos{\Theta}\ket{2} - \sin{\Theta}\ket{1}, \\
\tan{\Theta}\equiv \frac{\Delta}{2\cos{x}} -
\sqrt{\left(\frac{\Delta}{2\cos{x}}\right)^2 + 1}
\end{gathered}
\label{dressf}
\end{equation}
are eigenstates at a nonzero detuning. The corresponding
values of the quasienergy are
\begin{equation}
E_\Delta^{(\pm)} = \pm\sqrt{\frac{\Delta}{2}^2 + \cos^2{x}}.
\label{dresse}
\end{equation}
Figure \ref{fig7} shows a spatial variation of the quasienergies along the
standing-wave axis. It follows from Eqs.(\ref{dressf}) and (\ref{dresse})
that, in general case, atom moves in the two potentials $E_\Delta^{(\pm)}$
simultaneously.

At exact resonance, $\Delta=0$, the dressed states have the simple form
\begin{equation}
\ket{+} = \frac{1}{\sqrt 2}(\ket{1} + \ket{2}), \quad
\ket{-} = \frac{1}{\sqrt 2}(\ket{1} - \ket{2})
\label{dressd}
\end{equation}
and are called diabatic states. The resonant potentials,
$E_0^{(\pm)}=\pm\cos x$, cross each other at the nodes of the standing wave,
$x=\pi/2+\pi m$, $(m=0,\pm 1,\ldots)$. What will happen if we place the centroid
of an atomic wave packet exactly at the node, $x_0=\pi/2$, in the ground state
$\ket{1}$ and suppose its initial mean momentum to be zero, $p_0=0$?
The initial ground state is the superposition of the diabatic states:
$\ket{1}=(\ket{+}+\ket{-})/\sqrt2$. One part of the initial wave packet at the top of
the potential $E_0^{(+)}$ will start to move to the right under the action
of the gradient force $F^{(+)}=-dE_0^{(+)}/dx=\sin x$, and another one~--- to
the left to be forced by $F^{(-)}=-\sin x$ (see Fig.~\ref{fig7}). It is the
well-known optical Stern-Gerlach effect \cite{K75,Kaz,Sleator}. If the
maximal expected value of the atomic kinetic energy does not exceed the
potential one, the atom will be
trapped in the potential well. Two splitted components of the initial wave
packet will oscillate in the well with the period of oscillations
\begin{equation}
T\simeq 4 \sqrt{\frac{\pi}{\omega_r}}.
\label{period}
\end{equation}
The wave packet, with $p_0=0$, placed at the antinode, say, at $x_0=0$, is
simultaneously at the top of the potential $E_0^{(+)}$ and at the bottom of
$E_0^{(-)}$. Therefore, its $\ket{+}$-component will slide down the both sides
of the potential curve $E_0^{(+)}$, and the $\ket{-}$-component will oscillate
around $x=0$ (see Fig.~\ref{fig7}).

Out off resonance, $\Delta\ne0$, the atomic wave packet moves in the
bipotential $E_\Delta^{(\pm)}$~(\ref{dresse}). The distance between the
quasienergy curves is minimal at the nodes of the standing wave and equal
to $\Delta$ (see Fig.~\ref{fig7}). The spatial period and the modulation
depth of the resonant potentials $E_0^{(\pm)}$ are twice as much as those
for the nonresonant potentials $E_\Delta^{(\pm)}$.

The probability of nonadiabatic transitions between the dressed states
$\ket{+}_\Delta$ and $\ket{-}_\Delta$ can be estimated in a simple way.
The time of flight over a short distance $\delta x$ in neighbourhood of
a node is $\delta x/\omega_rp_\text{node}$. If the time of transition between
the quasienergy levels, $2/\Delta$, is of the order of the flight time, the
transition probability is close to $1$. It is easy to get  the characteristic
frequency of atomic motion from that condition~\cite{Kaz}
\begin{equation}
\Delta_0 = \sqrt{\omega_rp_\text{node}},
\label{freq}
\end{equation}
where $p_\text{node}$ is a value of the momentum in the vicinity of a node.

Depending on the relation between $\Delta$ and $\Delta_0$, there are three
typical cases.
\begin{enumerate}
\item If $|\Delta|\ll\Delta_0$, the nonadiabatic transition probability
between the states $\ket{+}_\Delta$ and $\ket{-}_\Delta$ upon crossing
any node is close to $1$. However, the diabatic states $\ket{+}$ and $\ket{-}$
are not mixed, and atom moves in one of optical resonant potentials.
\item If $|\Delta|\simeq\Delta_0$, the atom may or may not undergo a transition
upon crossing any node from one of the nonresonant potentials to another one
with the probabilities of the same order.
\item If $|\Delta|\gg\Delta_0$, the nonadiabatic transition probability
is exponentially small, and atom moves in one of the nonresonant potentials.
\end{enumerate}

\subsection{Wave packet motion in the momentum space}

The atom at $\tau=0$ is supposed to be prepared as a Gaussian wave packet
in the momentum space
\begin{equation}
a_0(p)=0, \quad
b_0(p)=\frac{1}{\sqrt{\sqrt{\pi}\Delta p}}
\exp\left[-\frac{(p-p_0)^2}{2(\Delta p)^2} -i(p-p_0)x_0\right],
\label{gaussp}
\end{equation}
with the momentum width $\Delta p=10$ corresponding to the
spatial width $\Delta X=\lambda_f/40\pi$ that is much smaller than the
optical wavelength $\lambda_f$. We compute the probability to find
a two-level atom at the moment of time $\tau$ with the momentum $p$
\begin{equation}
W(p,\tau)= |a(p,\tau)|^2 + |b(p,\tau)|^2,
\label{probp}
\end{equation}
by integrating Eqs.(\ref{Schp}) with the initial condition (\ref{gaussp}).
The recoil frequency, $\omega_r=10^{-5}$, is fixed and the centroid  of the
wave packet is placed at the antinode $x_0=0$, in all the numerical
experiments.

\subsubsection{Adiabatic evolution at exact resonance}
\begin{figure}
\begin{center}
\includegraphics[width=0.48\textwidth,clip]{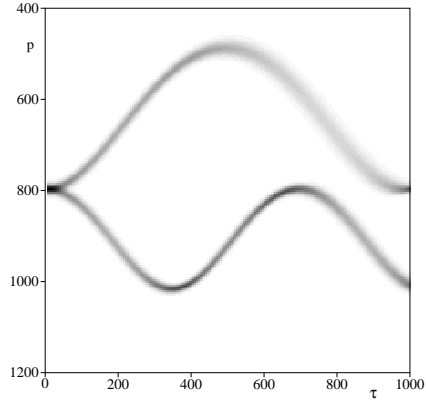}
\end{center}
\caption{Time dependence of the momentum probability function $W(p,\tau)$
for a ballistic atom at resonance prepared initially in the ground state
($\Delta=0$, $\omega_r=10^{-5}$, $x_0=0$, $p_0=800$).
}
\label{fig8}
\end{figure}
At exact resonance, $\Delta=0$, the wave functions of the diabatic states
$\ket{+}$ and $\ket{-}$ evolve independently, each one evolves in its own potential
$E_0^{(+)}$ and $E_0^{(-)}$, respectively. The atom, prepared initially in
the ground state $\ket{1}=(\ket{+}+\ket{-})/\sqrt2$ with the mean initial momentum
$p_0=800$, will start to move from the top of $E_0^{(+)}$ and the bottom of
$E_0^{(-)}$ potentials (see Fig.\ref{fig7}). Thus, the initial wave packet
will split into two components $\ket{+}$ and $\ket{-}$. Time evolution of the
probability function (\ref{probp}) for each of the components is shown in
Fig.\ref{fig8}. Pay, please, attention that the values of $p$ on this and similar
plots increase downwards. Color in this figure codes the values of $W(p,\tau)$.
The $\ket{+}$-component (the lower trajectory in the figure) slides down the
curve $E_0^{(+)}$ and, therefore, moves with an increasing velocity up to
the next antinode at $x=\pi$, and then it slows down approaching the antinode
at $x=2\pi$. The atom moves in the positive direction of the axis $x$ and
the process repeats periodically with the period
$\tau^{(+)}_0=2\pi/\omega_r\bar p^{(+)}_{0,2\pi}\simeq690$,
where $\bar p^{(+)}_{0,2\pi}$ is a mean
momentum of the $\ket{+}$-component upon the atomic motion between $0$ and
$2\pi$.

The $\ket{-}$-component (the upper trajectory in Fig.\ref{fig8}) moves upward the
potential curve $E_0^{(-)}$ and slows down up to reaching the top of
$E_0^{(-)}$ at $x=\pi$. Then it moves with an increasing momentum up to
$x=2\pi$. Since the mean momentum of the $\ket{-}$-component is smaller than
that of the $\ket{+}$ one, the corresponding period is longer, $\tau^{(-)}_0
\simeq 980$.

\subsubsection{Proliferation of wave packets at the nodes of the standing wave}
\begin{figure}
\begin{center}
\includegraphics[width=0.48\textwidth,clip]{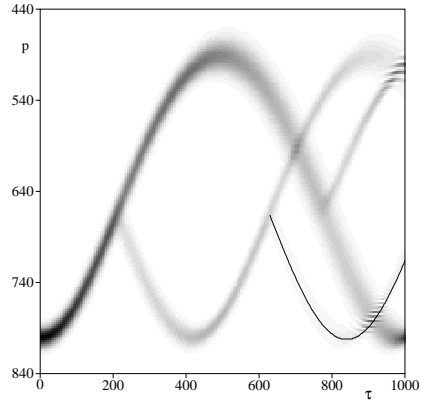}
\end{center}
\caption{Proliferation of atomic wave packets at the nodes of the
standing wave at the detuning $\Delta=0.05$. The atom is prepared
initially in the dressed state $\ket{+}$. Other conditions are the same as in
Fig.\ref{fig8}.
}
\label{fig9}
\end{figure}
New features in propagation of atomic wave packets through the standing wave
appear under the condition $\Delta\simeq\Delta_0$.
Using the semiclassical expression for the total atomic energy (\ref{H}),
let us estimate the value of the atomic
momentum at the nodes of the standing wave if the detuning is not large,
$|\Delta|\ll1$. If the atom is prepared initially in the state $\ket{+}$, i.e.,
$u_0=1$, $z_0=0$, and $x_0=0$ then we have $H=H_0=2.2$ at $p_0=800$. Since
the total energy is a constant, we get immediately from Eq.~(\ref{H})
\begin{equation}
p_\text{node} \simeq \sqrt{2H/\omega_r} \simeq 665.
\label{pnode}
\end{equation}
Using the same formula (\ref{H}), we get the values of the minimal
and maximal momenta if the atom starts to move with the initial mean momentum
 $p_0=800$:
$p_\text{min}\simeq\sqrt{2(H_0-1)/\omega_r}\simeq 490$ and
$p_\text{max}\simeq\sqrt{2(H_0+1)/\omega_r}\simeq 800$.

The formula (\ref{freq}) gives us the value of the characteristic frequency
under the chosen conditions, $\Delta_0\simeq 0.08$. We fix $\Delta=0.05$
in this section, so $\Delta\simeq\Delta_0$. The initial state $\ket{+}$ is
the following superposition of the adiabatic states:
\begin{equation}
\ket{+} = \frac{1}{\sqrt{2}}[(\cos{\Theta} + \sin{\Theta})\ket{+}_{\Delta} + (\cos{\Theta} - \sin{\Theta})\ket{-}_{\Delta}].
\label{dress+}
\end{equation}
With the help of (\ref{dresse}) we can estimate the mixing angle at
$\Delta=0.05$ to be equal to $\theta\simeq-\pi/4$. Then it follows from
(\ref{dress+}) that $\ket{+}\simeq\ket{-}_\Delta$, i.~e., practically all the wave
packet is initially at the bottom of the potential $E_\Delta^{(-)}$
(Fig.~\ref{fig7}). Figure~\ref{fig9} demonstrates that the wave packet really
slows down, and its centroid  intersects the node $x=\pi/2$ at
$\tau_1^{(-)}\simeq215$. Under the condition $\Delta\simeq\Delta_0$, the atom
has a probability to change the potential for another one upon crossing
a node and a probability to stay in its present potential.
This is exactly what we see in fig.~\ref{fig9}: the wave
packet splits at the node $x=\pi/2$ with the $\ket{-}$-component climbing over
the potential $E_\Delta^{(-)}$ (see the upper trajectory in this figure) and
the $\ket{+}$-component sliding down the curve $E_\Delta^{(+)}$ with an increasing
momentum (see the lower trajectory). Just after crossing the node,
the most part of the probability density
moves in the potential $E_\Delta^{(-)}$ because
the corresponding probability is larger. The $\ket{+}$-component increases its
velocity upon approaching the antinode at $x=\pi$ and then slows down up to
the second node at $x=3\pi/2$ where it splits into two components at
$\tau_2^{(+)}\simeq640$. After that, one of the components will move in the
potential $E_\Delta^{(+)}$ decreasing the velocity up to the next antinode
at $x=2\pi$, and the other one will move in $E_\Delta^{(-)}$ increasing its
velocity in the same space interval. The probability density of this
$\ket{-}$-component is only a few percents, and we draw a solid curve along this
trajectory in order to visualize the motion.
\begin{figure}
\begin{center}
\includegraphics[width=0.48\textwidth,clip]{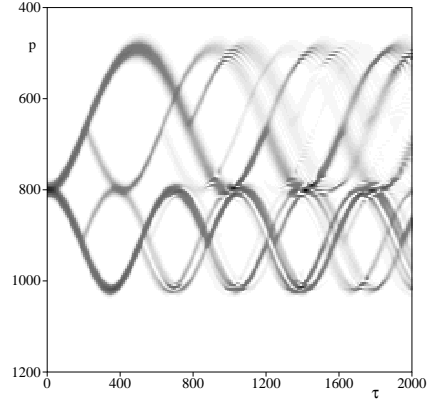}
\end{center}
\caption{The same as in Fig.\ref{fig9} but for the atom prepared initially
in the ground state.}
\label{fig10}
\end{figure}

The $\ket{-}$-component of the packet, splitted after crossing the first node at
$x=\pi/2$, has a smaller mean momentum than the $\ket{+}$-one. Therefore, it
reaches the second node later, at $\tau_2^{(-)}\simeq800$, where it splits
into two parts: the upper $\ket{+}$-component will move in the potential
$E_\Delta^{(+)}$ and the lower $\ket{-}$-one~--- in $E_\Delta^{(-)}$. Such a
proliferation of atomic wave packets takes places upon crossing all the next
nodes of the standing wave.

The moment of time $\tau_n^{(\pm)}$, when the centroids of the
$\ket{\pm}$-components cross the $n$-th node, can be estimated by the simple
formula (we suppose that the centroid of the atomic wave packet was at $x=0$ at
$\tau=0$):
\begin{equation}
\omega_r\overline{p}_{n-1,n}^{(\pm)}\tau_n^{(\pm)}=(2n-1)\frac\pi2, \quad
n=2,3, \dots.
\label{tnode}
\end{equation}
where $\overline{p}_{n-1,n}^{(\pm)}$ is a mean momentum of the
$\ket{\pm}$-components upon their movement between  $(n-1)$-th and $n$-th nodes.
This quantity for the $\ket{-}$-component, moving between $x=0$ and $x=\pi/2$, is
$\bar p_{0,1}^{(-)}=(p_0+p_\text{node})/2\simeq732.5$. So, the centroid of this wave
packet crosses the first node at $\tau_1^{(-)}\simeq214$. The lower
$\ket{+}$-component crosses the second node at $x=3\pi/2$ at
$\tau_2^{(+)}\simeq642$. For the upper $\ket{-}$-component we get
$\bar p_{1,2}^{(-)}=(p_\text{node}+p_\text{min})/2\simeq577.5$ and $\tau_2^{(-)}\simeq815$.
All the other moments of time, $\tau_n^{(\pm)}$, can be estimated in the
same way. The estimates obtained fit well  the numerical data
(see Fig.\ref{fig9}). The interference fringes on the upper trajectory at
$\tau\simeq1000$ and $p\simeq500$ and on the lower one at $\tau\simeq900$
and $p\simeq800$ reflect the fine-scale splitting of the corresponding
wave packets.

Let us now compute the probability map for the atom prepared initially
in the ground state $\ket{1}$ which has the following form in the adiabatic
state basis:
\begin{equation}
\ket{1} = \cos{\Theta} \ket{+}_\Delta - \sin{\Theta}\ket{-}_\Delta,
\label{dress1}
\end{equation}
It follows from (\ref{dresse}) that (\ref{dress1}) is almost a
$50\%$--$50\%$ superposition of the $\ket{+}_\Delta$ and $\ket{-}_\Delta$ states.
All the other conditions are assumed to be the same as before. The atomic
wave packet splits from the beginning into two components with the
$\ket{+}$-one sliding down the curve $E_\Delta^{(+)}$ (the lower trajectory
in Fig.~\ref{fig10}) and the $\ket{-}$-one climbing over the potential
$E_\Delta^{(-)}$ (the upper trajectory). Each of the components splits
at the first node with a small time difference between the events. The
subsequent proliferation of the wave packets occurs for the upper and lower
parts of the probability density independently on each other in
accordance with the same scenario as described above. In difference from the
preceding case, the atom, prepared initially in the ground state, acquired the values
of the momentum that are larger then the initial momentum $p_0=800$.
\begin{figure}
\begin{center}
\includegraphics[width=0.48\textwidth,clip]{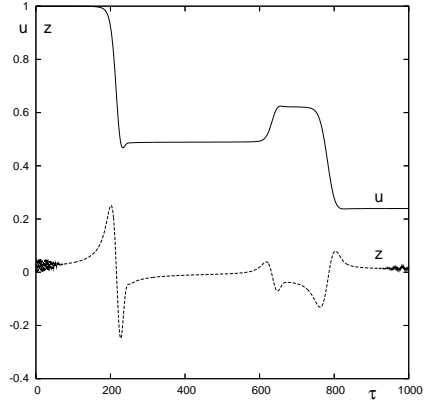}
\end{center}
\caption{Time dependence of the dipole moment $u$ and the population inversion
$z$ at the same conditions as in Fig.~\ref{fig9}.
}
\label{fig11}
\end{figure}

The nonadiabatic transitions are accompanied by drastic changes
in the internal state of the atom which is characterized by the values of
the synphased component of the electric dipole moment $u$ and the
population inversion $z$. In Fig.~\ref{fig11} we demonstrate their behavior
for the atom prepared initially in the state $\ket{+}$. Both the variables
change their values abruptly in the time intervals with the centers at
$\tau\simeq215$, $640$ and $815$, i.~e., when the centroids  of the atomic wave
packets cross the first two nodes.

\subsubsection{Adiabatic motion at large detunings}

\begin{figure}
\begin{center}
\includegraphics[width=0.48\textwidth,clip]{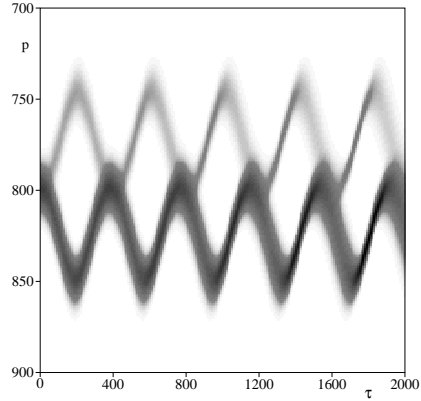}
\end{center}
\caption{Adiabatic evolution of the momentum probability function $W(p,\tau)$
for a ballistic atom at the large detuning $\Delta=2$.
}
\label{fig12}
\end{figure}

\begin{figure}
\begin{center}
\includegraphics[width=0.48\textwidth,clip]{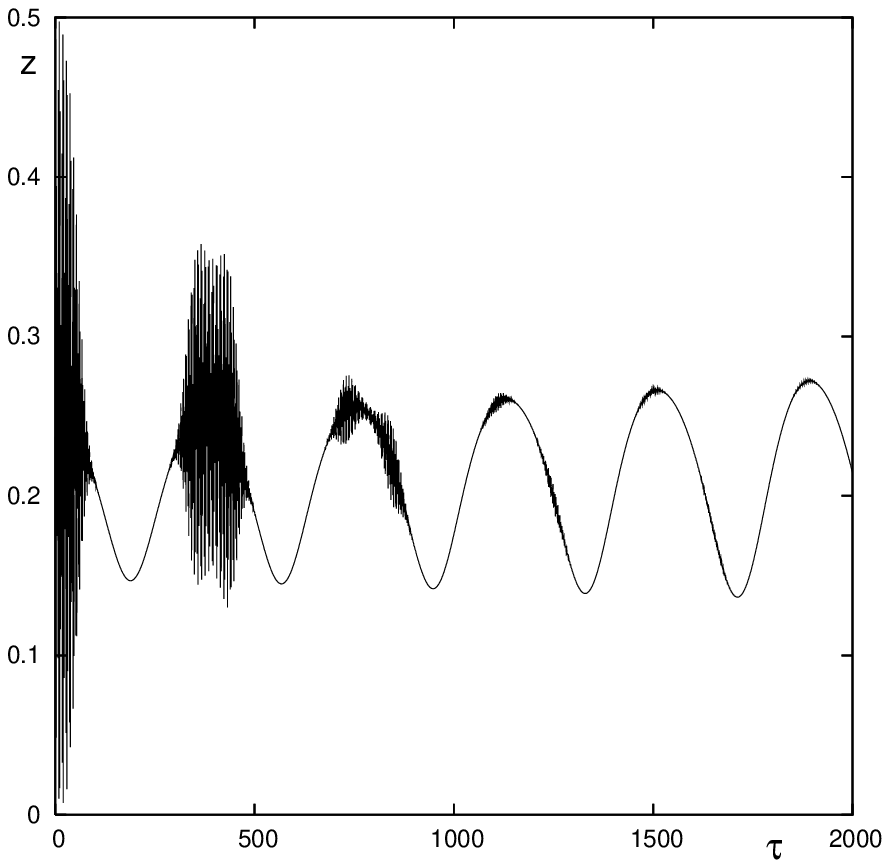}
\includegraphics[width=0.48\textwidth,clip]{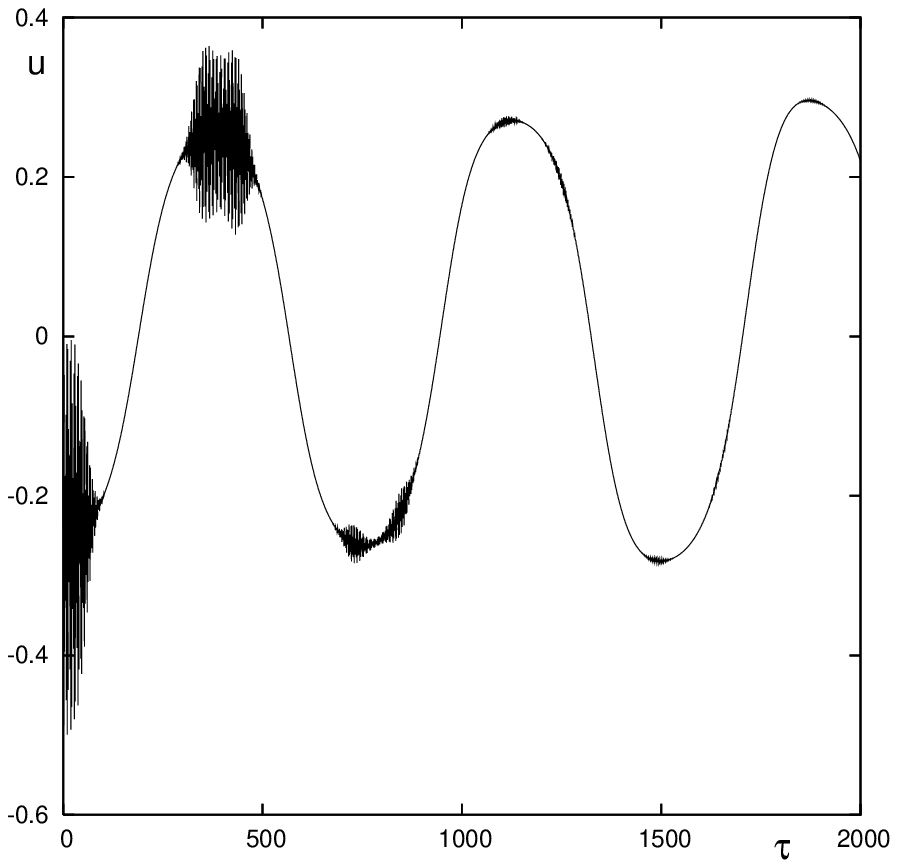}
\end{center}
\caption{The same as in Fig.\ref{fig11} but at the large detuning $\Delta=2$.
}
\label{fig13}
\end{figure}

For comparison with the results of the preceding section, we demonstrate in
Fig.~\ref{fig12} the evolution of the momentum distribution function
$W(p,\tau)$ with the ground initial state at $\Delta=2$ and the other same conditions as in the
preceding section. The detuning $\Delta=2$ is large as compared to the characteristic
frequency $\Delta_0\simeq0.09$ that is estimated from (\ref{freq}) at
$p_0=800$. It follows from (\ref{dressf}) and (\ref{dresse}) that
at $\Delta=2$ the initial state $\ket{1}$ is a superposition of approximately
$70\%$ of the state $\ket{+}_\Delta$ and $\sim30\%$ of the state
$\ket{-}_\Delta$. So
the main part  of the initial packet begins to move in the potential $E_\Delta^{(+)}$
increasing the momentum upon approaching the node at $x=\pi/2$, and the
other part moves in $E_\Delta^{(-)}$ decreasing the momentum in the same
space interval (see Fig.~\ref{fig12}). Upon crossing the nodes, the probability
of transition between the states $\ket{\pm}_\Delta$ is small if
$|\Delta|\gg\Delta_0$, and each of the component will continue to move in
its own potential. The process is repeated and we see the periodic variations
of the mean momentum of each of the components.  The same picture is observed
if we take the state $\ket{+}=(\ket{1}+\ket{2})/\sqrt2$ as the initial one. At
$\Delta=2$, the state $\ket{+}$ is a mix of $70\%$ of $\ket{-}_\Delta$
and $30\%$ of $\ket{+}_\Delta$, so the main part of the initial $\ket{+}$
wave packet will move in the potential $E_\Delta^{(-)}$. The evolution of
the internal atomic variables $z$ and $u$ is shown in Fig.~\ref{fig13}. There are
no jumps of $z$ and $u$ when the atom crosses nodes.  Instead of that, we see
fast oscillations of those variables when the atom crosses the first antinodes.

Thus, at $|\Delta|\gg\Delta_0$, there are no nonadiabatic transitions due to the
corresponding small probability and, therefore, no proliferation of wave packets
at the nodes. The evolution of the atomic wave packet is adiabatic.

\subsubsection{An atom can fly and be trapped simultaneously}

An intriguing effect of simultaneous trapping of an atom in a well of the
optical potential and its ballistic flight through the optical lattice is
observed at comparatively small values of the detuning. Let us prepare an atom in the
ground state $\ket{1}$ with such a mean initial value of the momentum $p_0$
that its $\ket{-}$-component would not be able to overcome the barrier of the
potential $E_\Delta^{(-)}$ but its $\ket{+}$-component would have a sufficient
kinetic energy to overcome the barrier of the $E_\Delta^{(+)}$ potential.
Now one could
expect periodic oscillations in the first well of the potential
$E_\Delta^{(-)}$ and a simultaneous ballistic flight in the $E_\Delta^{(+)}$
potential with a proliferation of wave packets of the $\ket{+}$-component at the nodes
of the standing wave.

\begin{figure}
\begin{center}
\includegraphics[width=0.48\textwidth,clip]{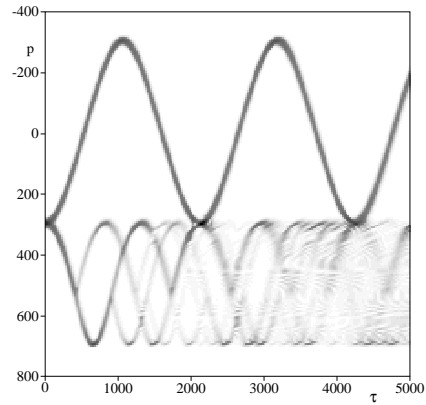}
\end{center}
\caption{Effect of simultaneous trapping of an atom in a well of the optical
potential and its flight through the wave. The ground initial state,
$\Delta=-0.05$, $p_0=300$.}
\label{fig14}
\end{figure}

Figure \ref{fig14} demonstrates this effect at $p_0=300$, $\Delta=-0.05$ and
the same other conditions as before. We see that the momentum of the
$\ket{-}$-component (the upper trajectory in this figure) oscillates in the range
($300$, $-300$), and this component is trapped in the first well
($-\pi/2\le x\le\pi/2$). Whereas the $\ket{+}$-component moves in the positive
direction splitting at each node. Estimates of the period of oscillations
of the $\ket{-}$-component, $T\simeq2240$, with the help of (\ref{period}) and
of the time when the centroid  of the $\ket{+}$-component crosses the first node,
$\tau_1^{(+)}\simeq380$ (formula (\ref{tnode})), fit well the data
in Fig.~\ref{fig14}.

\section{Quantum-classical correspondence and manifestations of dynamical
chaos in wave-packet atomic motion}

Dynamical chaos in classical systems is characterized by exponentially fast
divergence of initially close trajectories in a bounded phase space.
Such a behavior is possible because of the continuity of the classical
phase space whose points (therefore, classical system's states) can be
arbitrary close to each other. The trajectory concept is absent in quantum
mechanics whose phase space is not continuous due to the Heisenberg uncertainty
relation. The evolution of an isolated quantum system is unitary, and there
can be
no chaos in the sense of exponential sensitivity of its states to small variations in
initial conditions. What is usually understand under
``quantum chaos'' is special features of the unitary evolution of a quantum
system in the range of its parameter values and initial conditions at which
its classical analogue is chaotic.

The question ``what happens to classical motion in the quantum world'' is
a core of the problem of quantum-classical correspondence. In spite of
years of discussions from the beginning of the quantum era, it is still
unclear how classical features appear from the underlying quantum equations.
It is especially difficult to specify what happens to classical dynamical
chaos in the quantum world \cite{BZ78,Casati79,Z81,Gutzwiller,Reichl,Haake,
Shtokman}. The interest to the problem of ``quantum chaos'' is motivated by our
desire to understand the quantum origin of the observed classical chaos.

In this section we establish a correspondence between the quantized motion
of a two-level atom in a standing laser wave and its semiclassical analogue
considered in the third section. Semiclassical equations (\ref{mainsys})
represent a nonlinear dynamical system with positive
values of the maximal Lyapunov exponent in a wide range of the initial
conditions and control parameters $\omega_r$ and $\Delta$.
In other words, trajectories in the five-dimensional phase space are
exponentially sensitive to small variations in initial conditions and/or
parameters in those ranges. That local dynamical instability is a reason for
chaotic Rabi oscillations and chaotic motion of the atomic center of mass
discussed in the third section. In particular, it has been found that an
atom is able to walk chaotically in a strictly periodic optical lattice without any
noise or other random processes (see Fig.~\ref{fig2}). The chaotic behavior
is caused by jumps
of the electric-dipole moment $u$ at the nodes of the standing wave
(Fig.~\ref{fig3}). It follows from Eqs. (\ref{mainsys}) that this
quantity governs the atomic momentum. A stochastic map for the quantity $u$
(\ref{u_m}) allowed to derive analytic expressions for probability density
functions of the atomic trapping and
flight events that have been shown to fit well numerical simulation
\cite{PRA07}.

It has been shown that sudden changes in the behavior of $u$ take place
when we quantized the atomic motion (see Fig.~\ref{fig11}) under the condition
$\Delta\simeq\Delta_0$. Those changes are more smooth than the jumps of $u$ in
the semiclassical case because a delocalized wave packet crosses a node for a
finite time interval. The quantum analysis provides a clear reason for those
jumps at $\Delta\simeq\Delta_0$, namely, it is nonadiabatic transitions
between the quasienergy states $\ket{+}_\Delta$ and $\ket{-}_\Delta$ which occur
when an atom crosses any node of the standing wave. Those jumps are accompanied
 by splitting of
wave packets at the nodes. We may conclude that the proliferation of wave
packets at the nodes of the standing wave is a manifestation of classical
chaotic transport of an atom in an optical lattice that has been shown in
Refs.~\cite{JETP03,JRLR06,PRA07} to take place in exactly the same ranges of
initial conditions and control parameters. In particular, the effect of
simultaneous trapping of an atom in a well of the optical potential and its
flight in the same potential (Fig.~\ref{fig14}) is a quantum analogue of a chaotic walking of
an atom shown in Fig.~\ref{fig2}.

In conclusion we would like to discuss briefly the role of
dissipation. We did not take into account any losses in our treatment.
Coherent evolution of the atomic
state in a near-resonant standing-wave laser field is interrupted
by spontaneous emission events at random moments of times. The semiclassical
Hamiltonian evolution between these events has been shown to be
regular or chaotic depending on the values of the detuning $\Delta$ and the initial momentum $p_0$.
We stress that dynamical chaos may happen without any noise and any modulation
of the lattice parameters. It is a specific kind of
dynamical instability in the fundamental interaction
between the matter and radiation.

Dissipative transport of spontaneously emitting atoms in a 1D standing-wave
laser field has been studied in detail in Ref.~\cite{PRA08} in the regimes
where the underlying semiclassical Hamiltonian
dynamics is regular and chaotic. A Monte Carlo stochastic wavefunction method
was applied to simulate semiclassically the atomic
dynamics with coupled internal and translational degrees of
freedom. It has been shown in numerical experiments and confirmed
analytically that chaotic atomic transport can take the form either of
ballistic motion or a random walking with specific statistical properties.
The character of spatial and momentum
diffusion in the ballistic atomic transport was shown to change abruptly in the atom-laser detuning
regime where the Hamiltonian dynamics is irregular in
the sense of dynamical chaos. A clear correlation
between the behavior of the momentum diffusion coefficient
and Hamiltonian chaos probability  has been found.

What one could expect if spontaneous emission would be taken into
consideration with our fully quantum equations of motion? Any act of
spontaneous emission interrupts a coherent evolution of an atom
at a random  time
moment and is accompanied by a momentum recoil and a sudden transition of
the atom into the ground state which is a superposition of the dressed states.
The coherent evolution starts again after that. A collapse of the atomic wave
function and a splitting of atomic wave packets are expected just after
any spontaneous emission event. That additional splitting of wave packets
at random  time moments, besides of their proliferation at the nodes of
a standing wave at $\Delta\simeq\Delta_0$, can improve  the
quantum-classical correspondence in the regime of Hamiltonian chaos.

I thank L. Konkov and M. Uleysky for their help in preparing some figures.
This work was supported  by the Russian Foundation for Basic Research
(project no. 09-02-00358) and by the Program
``Fundamental Problems of  Nonlinear Dynamics'' of the Russian
Academy of Sciences.

\end{document}